\documentclass{article}
\usepackage{graphicx}
\graphicspath{%
    {converted_graphics/}
    {/}
}
\begin{document}

\begin{center}
{\LARGE Managing Systematic Errors}\vskip6pt

{\LARGE in Ice Crystal Growth Experiments}\vskip6pt

{\Large Kenneth G. Libbrecht}\vskip4pt

{\large Department of Physics, California Institute of Technology}\vskip-1pt

{\large Pasadena, California 91125}\vskip-1pt

\vskip18pt

\hrule\vskip1pt \hrule\vskip14pt
\end{center}

\textbf{Abstract}: We describe how surface interactions can affect the
growth of ice crystal facets in contact with a substrate by lowering the
normal nucleation barrier on the ice surface. We also describe how the
resulting enhanced growth rates can produce systematic errors even when
measuring the growth of facets not contacting the substrate. From an
analysis of the diffusion dynamics we then develop a simple procedure for
approximately correcting ice growth data, thus substantially reducing these
systematic errors. We have found this technique to be quite useful for
interpreting ice growth data and extracting the intrinsic attachment
coefficients of ice surfaces.

\section{Introduction}

We recently developed a new experimental apparatus for making precise
measurements of the growth rates of ice facet surfaces as a function of
temperature and supersaturation \cite{vig1, vig2}. During the course of our
investigations, we found that interactions with our sapphire substrate
introduced systematic errors that could be substantial in some
circumstances. In particular, we found that ice facets that were in contact
with the substrate tended to grow faster than otherwise expected, in a
variable and somewhat unpredictable fashion. This enhanced growth from
substrate interactions was particularly evident at low supersaturations.

\begin{figure}[ht] 
  \centering
  \includegraphics[width=5.6in,keepaspectratio]{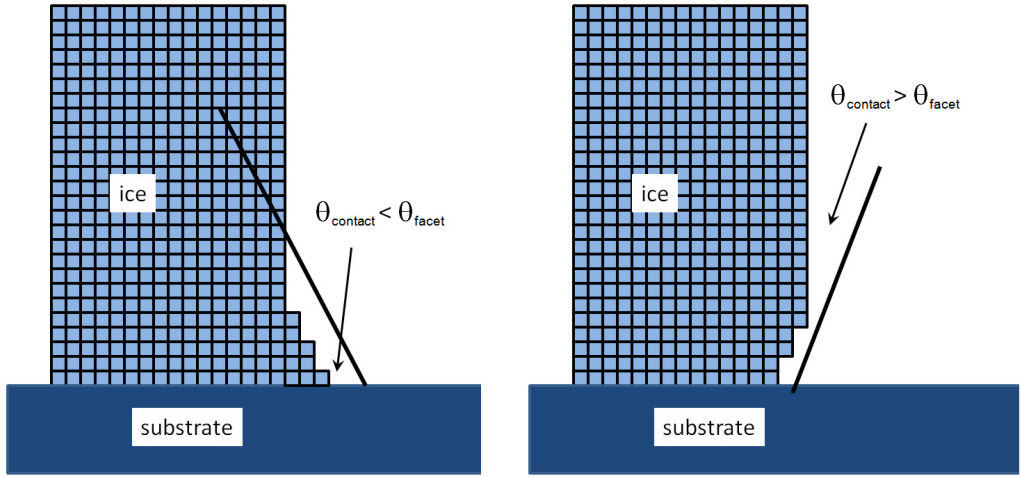}
  \caption{An ice crystal resting on a
substrate with $\protect\theta _{facet}=90$ degrees, which is the case when
the basal facet of a hexagonal prism rests on the substrate. At left $%
\protect\theta _{contact}<$ $\protect\theta _{facet},$ so substrate
interactions introduce molecular steps on the intersecting facet surface,
thus reducing the nucleation barrier that normally inhibits crystal growth
on that surface. At right $\protect\theta _{contact}>\protect\theta _{facet}$
and the nucleation barrier is not affected.}
  \label{contactangles}
\end{figure}

We believe that the enhanced growth arises because the ice/substrate
interaction creates molecular steps on the intersecting ice surface, thus
lowering the normal nucleation barrier that is present on an ice facet. A
contact-angle model is sufficient to explain these observations, as
illustrated in Figure \ref{contactangles}. If the ice/substrate contact
angle $\theta _{contact}$ is less than the intersection angle $\theta
_{facet}$ between the facet and the substrate, then the ice/substrate
contact will act as a source of molecular steps on the ice facet, as seen in
the left diagram in Figure \ref{contactangles}. The presence of these
molecular steps then reduces the normal nucleation barrier on the ice facet,
and the growth enhancement will be most pronounced at low supersaturation,
when the normal facet growth rate is especially low.

If $\theta _{contact}>\theta _{facet},$ however, then the contact will not
be a source of steps, and there will be no enhanced growth arising from
substrate interactions, as seen in the right diagram in Figure \ref%
{contactangles}. Since $\theta _{contact}$ depends on the chemical nature of
the substrate, as well as on any chemical residue on an imperfectly cleaned
substrate, we found that the ice growth behavior from substrate interactions
can be somewhat unpredictable.

This same effect appears to have affected other ice growth experiments
reported in the literature. For example, in \cite{becklac} the authors found
that prism facets not intersecting their substrate (facet type P in their
paper) grew approximately five times slower than facets that did intersect
the substrate, even while the facets were all exposed to essentially the
same growth conditions. Also, in \cite{nelsonknight} the authors found that
facets in contact with the exterior of their capillary tubes grew more
quickly than other facets, and that contact with the foreign surface also
reduced the measured nucleation barrier. Qualitatively the effects described
in these two experiments agreed with our observations, suggesting that all
three experiments were witnessing essentially the same phenomenon.

In addition, we found that substrate interactions can substantially
influence the growth of neighboring ice facets that are not directly in
contact with the substrate. The mechanism is that rapid growth of the
contacting facets lowers the supersaturation field around the crystal, owing
to particle diffusion through the background gas, and this reduces the
growth rates of the neighboring facets. This effect is very large for
background pressures near one bar, but even at 10-20 mbar (typical of our
measurements) it can introduce significant systematic errors. In addition,
as we see below, the effects can be important for measurements made both at
low and high supersaturations.

The remainder of this paper presents: 1) a detailed analysis of how
substrate interactions affect the growth rates of neighboring crystal
facets, those not in contact with the substrate; and 2) a method for
modeling and approximately correcting for substrate interactions.

\section{Systematic Errors from Substrate Interactions}

Our first, rather obvious, conclusion from these observations is that growth
measurements of ice facet surfaces in contact with a substrate may yield
little useful information on the intrinsic ice growth rates. A faceted
surface typically implies a nucleation barrier, and any reduction of that
barrier from substrate interactions likely produces unacceptably large
systematic errors. Of course, it is possible that substrate interactions are
negligible in some circumstances, but it rests on the experimenter to
convincingly demonstrate that this is the case.

We tried to find a suitable superhydrophobic coating for our sapphire
substrate that would increase $\theta _{contact}$ and thus reduce the
effects of substrate interactions, as shown in Figure \ref{contactangles}.
To date we have not found a surface treatment that is superior to a very
clean sapphire surface, but much research effort is currently being spent on
developing superhydrophobic coatings for a range of materials. We believe
that future measurements would likely benefit from finding suitable
superhydrophobic surface treatments.

Our second conclusion is that the indirect effects on neighboring facets can
be quite important also, even at low background pressures. Some diffusion
modeling is required to understand the magnitude of these effects on ice
crystal growth experiments. The remainder of this paper is focused on
outlining the diffusion modeling of this problem and developing a strategy
for managing and correcting the resulting systematic errors in experimental
data.

To develop a quantitative diffusion model, the present discussion considers
the problem of crystal growth in the presence of an inert background gas, in
which the transport of water vapor is governed by simple diffusion through
the gas background. We further simplify to the isothermal case, which is a
reasonably accurate approximation for growth on a substrate, since the
substrate acts as a thermal reservoir that readily absorbs the latent heat
generated from crystal growth \cite{libbrechtreview}.

\subsection{The Hemispherical Case with no Substrate Interactions}

To set the notation and basic scales in the problem, we first examine the
case of a hemispherical crystal of radius $R$ growing on an infinite
substrate. We assume a constant supersaturation $\sigma _{\infty }$ far from
the crystal and assume a single condensation coefficient $\alpha \left(
\sigma \right) $ for the crystal surface. The diffusion equation and
boundary conditions for this problem are equivalent to the spherical case,
which can be solved analytically. Following the notation in \cite%
{libbrechtreview}, the growth rate of the crystal surface is%
\begin{eqnarray*}
v &=&\alpha _{surf}v_{kin}\sigma _{surf} \\
&=&\frac{c_{sat}}{c_{solid}}D\frac{d\sigma }{dR}|_{surf}
\end{eqnarray*}%
where $\alpha _{surf}=\alpha (\sigma _{surf})$ is the condensation
coefficient at the ice surface and $\sigma _{surf}=\sigma \left( R\right) $.
The diffusion equation together with these mixed boundary conditions for $%
\sigma \left( r\right) $ give the solution \cite{libbrechtreview} 
\[
v=\frac{\alpha _{surf}\alpha _{diff}}{\alpha _{surf}+\alpha _{diff}}%
v_{kin}\sigma _{\infty } 
\]%
where 
\begin{eqnarray}
\alpha _{diff} &=&\frac{c_{sat}D}{c_{solid}v_{kin}R}=\frac{D}{R}\sqrt{\frac{%
2\pi m}{kT}}  \label{alphadiff} \\
&\approx &0.15\left( \frac{1\textrm{ }\mu \textrm{m}}{R}\right) \left( \frac{D}{%
D_{air}}\right)  \nonumber
\end{eqnarray}%
where $D$ is the diffusion constant in the background gas. For the final
expression we used $D_{air}=2\times 10^{-5}$ m$^{2}/\sec $ for the diffusion
constant of water vapor in air at a pressure of one bar. To a good
approximation $D$ is inversely proportional to the background gas pressure.

\begin{figure}[htb] 
  \centering
  \includegraphics[width=4.5in, keepaspectratio]{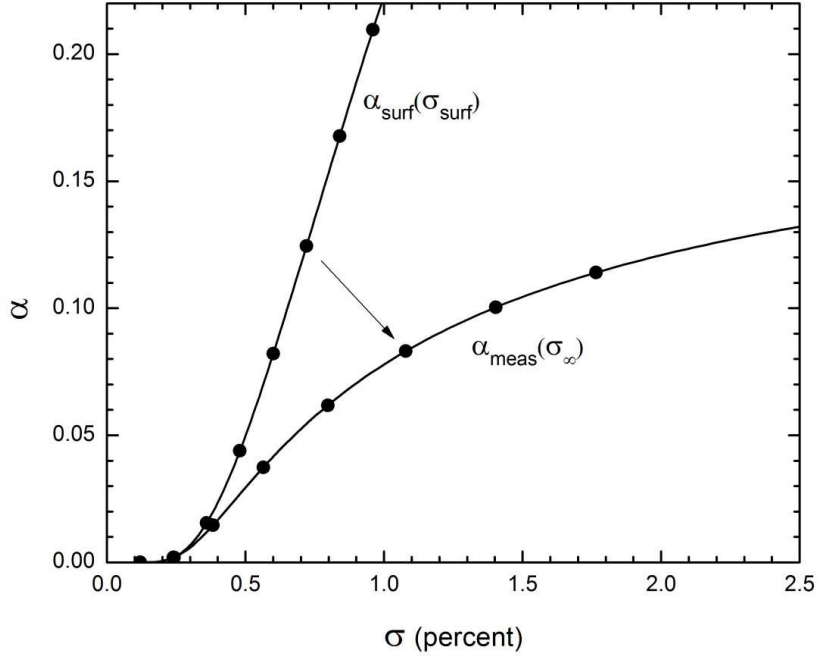}
  \caption{Calculated plots of the
functions $\protect\alpha _{surf}(\protect\sigma _{surf})$ and $\protect%
\alpha _{meas}\left( \protect\sigma _{\infty }\right) $ defined in the text,
for the case of a hemispherical crystal growing on a substrate. The
individual points show the correspondence between the two curves. Here we
used $\protect\alpha _{diff}=0.25,$ corresponding to a crystal radius of $%
R\approx 30$ $\protect\mu $m in a background pressure of 20 mbar (both
typical of our measurements), and an intrinsic attachment coefficient of $%
\protect\alpha _{surf}\left( \protect\sigma _{surf}\right) =\exp (-0.015/%
\protect\sigma _{surf})$ (typical for basal facet growth at temperatures
near -13 C) \protect\cite{vig2}.}
  \label{example1}
\end{figure}

From the analytic solution we also have the supersaturation at the ice
surface%
\begin{equation}
\sigma _{surf}=\frac{\alpha _{diff}}{\alpha _{surf}+\alpha _{diff}}\sigma
_{\infty }  \label{sigmaR}
\end{equation}%
and eliminating $\alpha _{surf}$ from this expression gives%
\begin{equation}
\sigma _{surf}=\sigma _{\infty }-\frac{v}{\alpha _{diff}v_{kin}}
\label{sigmaR2}
\end{equation}%
which is exact for the spherical and hemispherical cases.

When measuring ice crystal growth rates \cite{vig1, vig2}, we typically
establish a supersaturation $\sigma _{\infty }$ far from the crystal and
then measure the crystal dimensions as a function of time and $\sigma
_{\infty }$, with the goal of extracting the intrinsic attachment
coefficient $\alpha \left( \sigma _{surf}\right) $ for the growing surfaces.
In the limit of fast diffusion (or slow attachment kinetics), $\alpha
_{surf}\ll \alpha _{diff}$ and the hemispherical case gives%
\begin{equation}
\alpha _{surf}\approx \alpha _{meas}=\frac{v}{v_{kin}\sigma _{\infty }}
\label{alphameas}
\end{equation}%
which defines the measured quantity $\alpha _{meas}.$ If diffusion cannot be
neglected, then%
\begin{equation}
\alpha _{meas}=\frac{\alpha _{surf}\alpha _{diff}}{\alpha _{surf}+\alpha
_{diff}}  \label{alphameas2}
\end{equation}%
and a plot of $\alpha _{meas}\left( \sigma _{\infty }\right) $ typically
deviates from the desired $\alpha _{surf}\left( \sigma _{surf}\right) $ at
high $\sigma _{\infty }.$ An example of this is shown in Figure \ref%
{example1}, and plots of $\alpha _{meas}\left( \sigma _{\infty }\right) $
from our data show this same functional form \cite{vig1, vig2}.

The example in Figure \ref{example1} shows that $\alpha _{meas}\left( \sigma
_{\infty }\right) $ is approximately equal to $\alpha _{surf}\left( \sigma
_{surf}\right) $ at low supersaturations, which is precisely when $\alpha
_{surf}\ll \alpha _{diff}.$ Increasing $\sigma _{\infty }$ increases $\sigma
_{surf}$ until $\alpha _{surf}\left( \sigma _{surf}\right) $ becomes larger
and the criterion $\alpha _{surf}\ll \alpha _{diff}$ is no longer satisfied.
When this occurs, the two curves in Figure \ref{example1} deviate from one
another.

\subsubsection{Correcting the Hemispherical Case}

In our experiments we determine $\sigma _{\infty }$ and measure the growth
velocity $v$ of a facet not contacting the substrate, and from Equation \ref%
{alphameas} we derive $\alpha _{meas}\left( \sigma _{\infty }\right) $
entirely from measured quantities. Our goal, however, is to determine the
intrinsic condensation coefficient $\alpha _{surf}\left( \sigma
_{surf}\right) .$ For our idealized hemispherical crystal, we could apply a
correction to the data that transforms $\alpha _{meas}\left( \sigma _{\infty
}\right) $ into $\alpha _{surf}\left( \sigma _{surf}\right) $ using%
\begin{eqnarray}
\sigma _{surf} &=&\left( 1-\frac{\alpha _{meas}}{\alpha _{diff}}\right)
\sigma _{\infty }  \label{correction} \\
\alpha _{surf}\left( \sigma _{surf}\right)  &=&\left( 1-\frac{\alpha _{meas}%
}{\alpha _{diff}}\right) ^{-1}\alpha _{meas}  \nonumber
\end{eqnarray}%
where $\alpha _{diff}$ is derived from the measured crystal radius using
Equation \ref{alphadiff}. Clearly this correction works best when $\alpha
_{meas}\ll \alpha _{diff},$ or equivalently $\alpha _{surf}\ll \alpha
_{diff}.$ As long as the crystal growth is limited mainly by attachment
kinetics and not by diffusion, then we can use this correction to reasonably
extract $\alpha _{surf}\left( \sigma _{surf}\right) $ from the measured $%
\alpha _{meas}\left( \sigma _{\infty }\right) $.

\begin{figure}[ht] 
  \centering
  \includegraphics[width=4.5in,keepaspectratio]{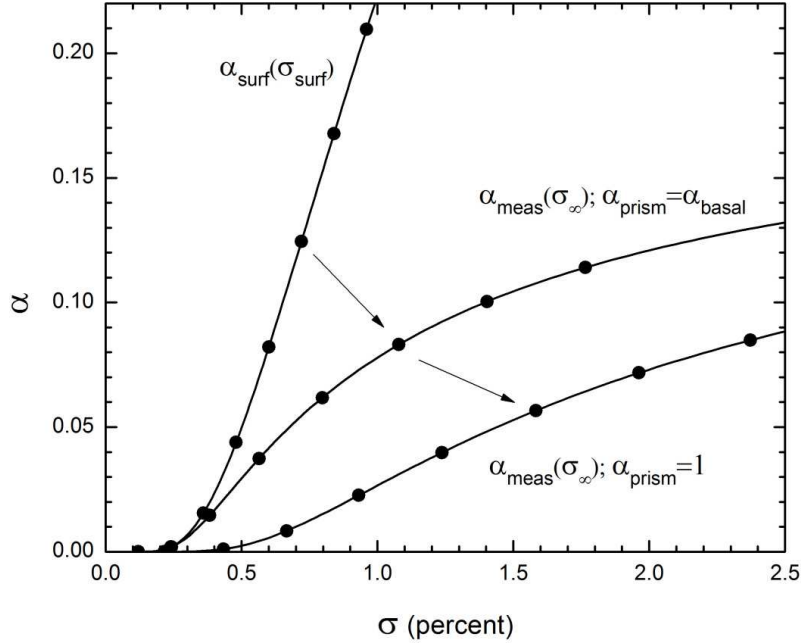}
  \caption{Calculated plots of the
functions $\protect\alpha _{surf}(\protect\sigma _{surf})$ and $\protect%
\alpha _{meas}\left( \protect\sigma _{\infty }\right) $ for growth of the
top basal surface of a hexagonal plate crystal, with the lower basal surface
resting on the substrate. The individual points show the correspondence
between the three curves. In this example we used the model of a circular
disk crystal of radius $R\approx 30$ $\protect\mu $m and thickness $H\approx
4$ $\protect\mu $m in a background pressure of 20 mbar, with the intrinsic
basal attachment coefficient $\protect\alpha _{basal}=\protect\alpha %
_{surf}\left( \protect\sigma _{surf}\right) =\exp (-0.015/\protect\sigma %
_{surf}).$ For the middle curve we used $\protect\alpha _{prism}=\protect%
\alpha _{basal},$ which gives $v_{prism}\approx v_{basal}$ and an $\protect%
\alpha _{meas}\left( \protect\sigma _{\infty }\right) $ curve that is
similar to that shown in Figure \protect\ref{example1}. For the lower curve
we used $\protect\alpha _{prism}=1$, giving $v_{prism}>v_{basal}$ and an $%
\protect\alpha _{meas}\left( \protect\sigma _{\infty }\right) $ curve that
is substantially distorted, especially at low $\protect\sigma _{\infty }.$}
  \label{example2}
\end{figure}

\subsection{Plate Growth with Substrate Interactions}

We next consider the growth of simple hexagonal plate crystals, which is
what we observed in \cite{vig2}. In this case one basal facet is resting on
the substrate, and we measure the growth of the opposite basal facet along
with the growth of the six prism facets. We typically have $v_{basal}\neq
v_{prism},$ and there is no analytic solution of the diffusion equation that
gives the supersaturation field $\sigma (x)$ around the crystal.

To see the problems that result, consider the example shown in Figure \ref%
{example2}, in which we examine the basal growth of a plate of radius $%
R\approx 30$ $\mu $m and thickness $H\approx 4$ $\mu $m, again at a pressure
of 20 mbar. If it so happens that $\alpha _{prism}(\sigma )=\alpha
_{basal}\left( \sigma \right) ,$ then $v_{prism}\approx v_{basal}$ and we
see that the resulting $\alpha _{meas}\left( \sigma _{\infty }\right) $
looks quite similar to the hemispherical case in Figure \ref{example1}, with
some $\alpha _{diff}$ appropriate for the crystal size and geometry. In
particular, we see that $\alpha _{meas}\left( \sigma _{\infty }\right)
\approx \alpha _{surf}\left( \sigma _{surf}\right) $ at low
supersaturations, as we saw with the hemispherical case.

If we increase the prism growth rate, however, in this example by setting $%
\alpha _{prism}=1,$ then the faster prism growth pulls down the
supersaturation around the entire crystal. Because $\alpha _{basal}\left(
\sigma \right) $ is a strong function of $\sigma ,$ this means that the
measured basal growth rates are substantially reduced. Furthermore, the
reduction factor is especially large at low supersaturations, as can be seen
by examining the curves in Figure \ref{example2}. Now this is a rather
extreme case, with exceptionally fast prism growth, but it demonstrates how
the measured $\alpha _{meas}\left( \sigma _{\infty }\right) $ can be
distorted at both high and low $\sigma _{\infty }$ by diffusion effects,
even at quite low background air pressures. These effects can make it
difficult to reliably determine $\alpha _{surf}\left( \sigma _{surf}\right) $
from the measurements, requiring diffusion modeling.

\subsubsection{Correcting Experimental Data}

In principle one could use numerical modeling of particle diffusion to
extract $\alpha _{surf}\left( \sigma _{surf}\right) $ from $\alpha
_{meas}\left( \sigma _{\infty }\right) ,$ provided the difference is not too
large. In practice, however, such modeling is quite computationally
intensive, and it is problematic to do a detailed numerical model for each
of hundreds of test crystals. We would therefore like a simpler analytic
approach that would allow us to apply a rough correction to our data. To
produce such an approximate correction, we generalize Equation \ref{sigmaR2}
by writing%
\begin{equation}
\sigma _{surf}\approx \sigma _{corr}=\sigma _{\infty }-\sum_{i}\frac{%
f_{i}v_{i}}{\alpha _{diff,eff}v_{kin}}  \label{sigcorr}
\end{equation}%
where the sum is over the different growing regions of the crystal (in this
case the one basal facet and six prism facets). The $f_{i}$ give the
fractional areas of the different regions (with $\sum f_{i}=1)$, where each
region is growing with velocity $v_{i}.$ From Equation \ref{alphadiff} we
take%
\[
\alpha _{diff,eff}\approx 0.15\left( \frac{1\textrm{ }\mu \textrm{m}}{R_{eff}}%
\right) \left( \frac{D}{D_{air}}\right) 
\]%
where%
\[
2\pi R_{eff}^{2}=\sum_{i}A_{i} 
\]%
and the $A_{i}$ give the areas of each crystal region. Note that Equation %
\ref{sigcorr} is exact for the hemispherical case, where it reduces to
Equation \ref{sigmaR2}. For the facet of interest we then have%
\[
\alpha _{surf}\left( \sigma _{surf}\right) \approx \frac{v}{v_{kin}\sigma
_{surf}}\approx \frac{\sigma _{\infty }}{\sigma _{corr}}\alpha _{meas} 
\]%
and this correction can be applied to each data point.

A key feature in this correction is that the $v_{i}$, $A_{i}$, and $f_{i}$
can all be extracted from the measured crystal size and geometry as a
function of time. If the prism growth is influenced by substrate
interactions, as we describe above, we can still use this formalism to
correct for changes in the supersaturation field that affect the basal
growth measurements. As long as the difference $\sigma _{\infty }-\sigma
_{corr}$ is not too great, we can determine $\alpha _{surf}\left( \sigma
_{surf}\right) $ for the basal surface.

\begin{figure}[tbp] 
  \centering
  \includegraphics[height=6.7in,keepaspectratio]{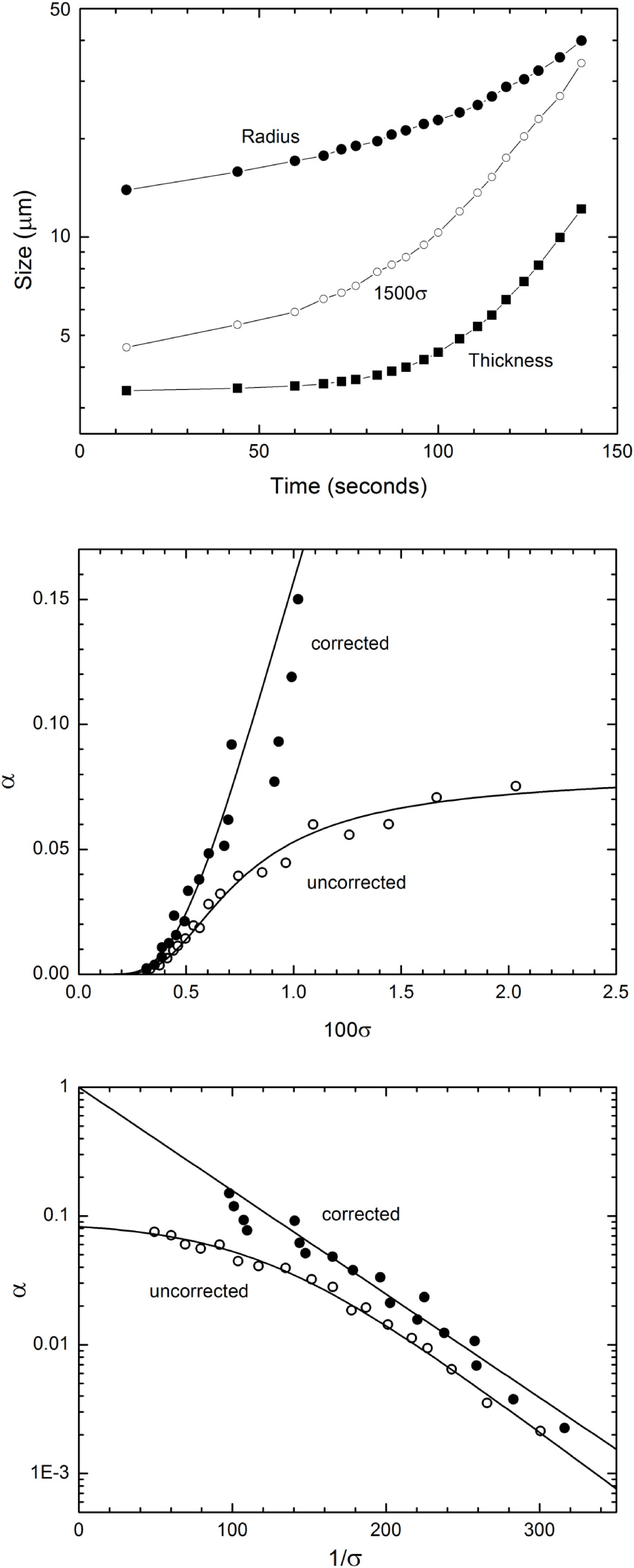}
  \caption{An example of actual ice
crystal growth data corrected using the procedure described in the text.
(Top) The measured radius and thickness of a hexagonal plate crystal as a
function of time during a growth run from \protect\cite{vig2}, along with
1500$\protect\sigma _{\infty }$ (in absolute units). The background pressure
during this measurement was 25 mbar. Note that at early times the radius
increased more rapidly than the thickness $(v_{prism}>v_{basal})$ because of
substrate interactions. (Middle, Bottom) The directly measured condensation
coefficient $\protect\alpha _{meas}(\protect\sigma _{\infty })$ (open
symbols) and the corrected measurement $\protect\alpha (\protect\sigma %
_{surf})$ (closed symbols). The fit lines in both these plots are $\protect%
\alpha _{meas}(\protect\sigma _{\infty })=\exp (-0.0205/\protect\sigma )%
\protect\alpha _{diff}/(\exp (-0.0205/\protect\sigma )+\protect\alpha %
_{diff}),$ with $\protect\alpha _{diff}=0.09,$ and $\protect\alpha (\protect%
\sigma _{surf})=\exp (-0.0185/\protect\sigma )$.}
  \label{combo}
\end{figure}

Figure \ref{combo} shows how this correction process works in practice,
using a plate-like crystal growing at a temperature of -12 C, taken from 
\cite{vig2}. Note that the prism growth rate at the beginning of this
measurement was $v_{prism}\approx 63$ nm/sec, when $\sigma _{\infty }\approx
0.33\%.$ If $\alpha _{prism}$ had been equal to unity, as we assumed in the
example in Figure \ref{example2}, then the radial growth would have been a
much faster $v_{prism}\approx 900$ nm/sec. Thus we verify that the $\alpha
_{prism}=1$ case in Figure \ref{example2} was somewhat extreme, and in our
actual data the corrections are usually substantially smaller.

Note also that our fit to the uncorrected data in Figure \ref{combo} gave $%
\alpha _{diff}=0.09$, a value roughly a factor of two smaller than the $%
\alpha _{diff}$ calculated from the size and geometry of the crystal. We
observed this discrepancy frequently in our basal growth data, and we find
it is nicely resolved by applying the above data correction. The corrected
data give a significantly better measure of the intrinsic attachment
coefficient $\alpha _{surf}\left( \sigma _{surf}\right) $, with reduced
systematic effects compared with using only the uncorrected $\alpha
_{meas}\left( \sigma _{\infty }\right) $. This improvement is accompanied by
only a modest increase in the scatter of the data points arising from the
imperfect correction process.

It is straightforward as well to apply this same correction process to
measurements of prism facet growth, where one prism facet rests on the
substrate. The area factors $f_{i}$ must be changed to fit the prism growth
case, but otherwise the prism growth correction is identical to the basal
case described above.

\section{Conclusion}

The approximate correction described above serves several purposes in the
analysis of our data: 1) it reduces the systematic errors in our
determination of $\alpha _{surf}\left( \sigma _{surf}\right) $ for the
growth of facets that do not contact the substrate; 2) it eliminates the fit
parameter $\alpha _{diff}$ when fitting the observed $\alpha _{meas}\left(
\sigma _{\infty }\right) $ (see \cite{vig2}); 3) a comparison of $\alpha
_{meas}\left( \sigma _{\infty }\right) $ and $\alpha _{surf}\left( \sigma
_{surf}\right) $ indicates which crystals have large correction factors and
thus should be given less weight in determining $\alpha _{surf}\left( \sigma
_{surf}\right) ;$ and 4) the correction allows us to make reliable
measurements of $\alpha _{surf}\left( \sigma _{surf}\right) $ over a broader
range of experimental conditions than we could without the data correction.

Although the correction is only approximately valid, it nevertheless does an
adequate job of reducing an important systematic effect. We have found it to
be a valuable tool for interpreting measurements of the growth of ice
crystals on substrates at low background pressures.

\end{document}